\documentstyle[sprocl]{article}

\def\Journal#1#2#3#4{{#1} {\bf #2}, #3 (#4)}

\def\NPB{{\em Nucl. Phys.} B}
\def\PLB{{\em Phys. Lett.}  B}

\def\be{\begin{equation}}
\def\ee{\end{equation}}
\def\bea{\begin{eqnarray}}
\def\eea{\end{eqnarray}}

\begin{document}

\title{
  \begin{flushright} {\small $\begin{array}{ l } \mbox{HD--THEP--99--05} \\
    \mbox{WUB--99--03} \end{array} $}
 \end{flushright}
\vspace{1cm}
ANALYTICAL STUDIES FOR THE CRITICAL LINE AND CRITICAL
ENDPOINT OF THE 
ELECTROWEAK PHASE TRANSITION\footnote{Talk presented by H.~Meyer-Ortmanns
at the {\it Strong and Electroweak Matter `98} Conference in
Copenhagen, December 2-5, 1998}
}

\author{H. MEYER-ORTMANNS}

\address{Fachbereich 8 Physik, Bergische Universit\"at Wuppertal,
Gau\ss str.20,\\ D-42097 Wuppertal, Germany \\
E-mail: ortmanns@theorie.physik.uni-wuppertal.de}

\author{T. REISZ}

\address{Institut f\"ur Theoretische Physik, Universit\"at Heidelberg,
Philosophenweg 16,\\
D-69120 Heidelberg, Germany\\E-mail: reisz@thphys.uni-heidelberg.de}


\maketitle\abstracts{ We outline two approaches for studying the 
electroweak phase transition in the framework of the four-dimensional
SU(2) Higgs model on a lattice. The first one is based on a combination
of variational estimates for the free energy and a solution of the resulting
gap equations by means of dynamical linked cluster expansions.
In the second approach we first indicate the derivation of 
an effective scalar field theory
upon integration over the gauge fields. The phase structure of the resulting
scalar theory is then investigated by means of generalized linked cluster
expansions. We present results for the critical line and the critical endpoint.
}

\section{Introduction}

We determine the critical line and critical endpoint
of the electroweak transition in the framework of the SU(2) Higgs model
on the lattice in four dimensions.
Usually the electroweak transition in the SU(2) Higgs model is studied with
Monte Carlo simulations either in the four-dimensional model
\cite{aoki}, 
\cite{cziskor}
or -after dimensional reduction- in an effective three-dimensional model
of the same form \cite{tsypin}. The time
extensions in such Monte Carlo simulations is restricted to small values
such as $N_\tau=2,3$ unless one uses asymmetric lattices, and $\lambda$
is of $O(10^{-4})$.
In our approach we first want to integrate out the gauge field degrees
of freedom and then study the phase structure of the resulting effective
scalar model with {\it Linked Cluster Expansions} (LCEs) that have been
developed to a powerful tool by extending the expansions 
to a high order in the 
expansion parameter and to a finite volume \cite{reisz1}, \cite{hilde1}.

\section{Critical Line from Variational Estimates and DLCEs}

In the first part we describe an analytic estimate of the critical line
$\kappa_c(\lambda)$ for a given gauge coupling $\beta$. The estimate is
based on a combination of variational estimates for the free energy of the
SU(2) Higgs model in 4D and {\it Dynamical Linked Cluster Expansions}
(DLCEs) for correlators in the set of variational equations. Already in
this first estimate the localization
of the critical line agrees quite well with high precision  
Monte Carlo results, since even this estimate became rather complex
after all.

Variational estimates for the free energy are based on the convexity
of the exp-function and the positivity of the measure. Under these
conditions we obtain an unequality between partition functions which
is of the form
\be
\exp{(-Vf)}\;\equiv \;Z\;\geq\;Z_{VE}\cdot \exp{<-(S-S_{VE}(\zeta))>_{VE}}
\; \equiv \; \exp{-V\widetilde{f}(\zeta)}
\ee
with the following notations. $V$ is the D-dimensional volume, $f$ is the
free energy density of the system described by the partition function $Z$,
in our case $Z$ is the partition function of the SU(2) Higgs model 
in D=4, $S_{VE}$ ($Z_{VE}$)
is the auxiliary action (partition function), respectively, that depends
on a generic set of variational parmeters $\zeta$, and $\widetilde{f}$ is
the trial free energy density which should be
optimized as a function of $\zeta$ in order to minimize 
$\widetilde{f}-f\geq 0$. The generic set of variational equations is then
given as
\be
     \partial_\zeta\widetilde{f}(\zeta) \; 
                       \vert_{\zeta=\widetilde{\zeta}} = 0 \qquad 
     \partial^2_\zeta\widetilde{f}(\zeta) 
                       \vert_{\zeta=\widetilde{\zeta}} >  0
\ee
An equality sign in the second unequality determines the critical 
temperature $T_c$
(coupling $\kappa_c$) in case of a second order transition and 
gives a lower
(upper) bound on $T_c$, ($\kappa_c$) in case of a first order transition.

Next we have to find a good choice for $S_{VE}$ in case of the SU(2) Higgs
model. 
Since we know  that the three-dimensional spatial part of the 
gauge-Higgs interaction is supposed to contain the non-perturbative 
degrees of freedom that drive the Higgs transition, in our
final choice for $S_{VE}$, we treat the physics along three-dimensional
spatial hypersurfaces as accurately as possible, but implement a factorization
along the time direction so that
\be\label{zve}
   Z_{VE}(\zeta_{link},\zeta_{cube},\xi)=Z^{L_0}_{cube}(\zeta_{cube},\xi)
                \cdot Z_{link}^{L_0V_3}(\zeta_{link})
\ee
in which $Z_{link}$ is an exactly solvable one-link partition function
depending on a variational parameter $\zeta_{link}$. $L_0$
is the extension in time direction, $V_3$ the 3-dimensional volume.
The action of the partition function $Z_{cube}$ for the 3-D 
hypersurfaces is given as $S_{cube}=S$ with
\be\label{S_cube}
  S=-\sum_{x\in\Lambda_3} 
   \biggl[ 2 \left( \xi {\rm Tr}\Phi(x)
          + \zeta_{cube} {\rm Tr}U(x;\mu) 
            \right) 
          +  \kappa \sum_{\mu=1}^{3}
               {\rm Tr}(\Phi^{\dagger}(x)U(x;\mu)\Phi(x+\mu)) 
   \biggr]
\ee
depending on variational parameters $\zeta_{cube}$ and $\xi$. Note
that in contrast to usual LCEs the hopping term in (\ref{S_cube}) does 
depend on $U$,
and $U$ has its own dynamics governed by the second term in $S_{cube}$.
Thus a Taylor expansion of $\ln{Z_{cube}}$ in powers of $\kappa$
about $\kappa=0$ will lead to a more general type of expansion, since the
familiar LCE only apply to frozen $U$-dynamics.
We will need such expansions for derived quantities from $\ln{Z_{cube}}$
that occur in the set of variational equations. The detailed form of the
variational equations is given in \cite{hilde2}. Here we only state that
they depend on connected n-point functions in $\Phi$s and $U$s 
up to $n=6$, evaluated w.r.t. $<\cdot>_{cube}$. It is these expectation
values that we evaluate as power series in $\kappa$ with DLCEs that 
have been introduced and systematized in \cite{hilde2}. The
number of graphs that contribute up to and including O($\kappa^4$) is
of the order of several hundred. Finally we solve the variational equations
for $\zeta_{link}$, $\zeta_{cube}$, $\xi$ and $\kappa_c$ as a power
series in $\kappa$. 

As results for $\kappa_c$ we obtain for example $\kappa_c=0.1282(1)$ 
($0.1281(1)$) for 
$\lambda=5\cdot 10^{-4}$ ($5.1\cdot 10^{-4}$), and 
$\beta=8$ as compared to $\kappa_c=0.12887(1)$ ($0.12852(2)$) from 
\cite{jansen}, respectively.
The good 
quantitative agreement of low order DLCEs (including O($\kappa^4$)) with
high precision Monte Carlo results appears less surprising in view of the
number of graphs that contribute to the series up to 
that order.

\section{Critical Endpoint of the Electroweak Phase Transition}

Starting from an SU(2)-Higgs model in $4D$ we have derived \cite{pinn} an
effective scalar model in the following steps. First we absorb the
angular
part of the U(2) Higgs field in a gauge transformation. The remaining
Higgs degrees of freedom are then given by scalar fields $\rho(x)$,
$\rho(x)\geq 0$, so that $Tr \Phi^\dagger\Phi=\rho^2$. Integrating
upon the $U$-dependent part of the original action leads to an
effective action $W(\rho)$. $W(\rho)$ is the free energy of the gauge
fields
in a scalar background field. Since we are interested
in an effective
model that describes the phase structure in the vicinity of the
critical endpoint  with long range correlations, we perform a small
momentum expansion of gauge field correlators 
about (lattice) momentum $\widehat{p}=0$ that appear as coefficients
in $W(\rho)$. It is justified if
$p/m_{glue}<<1$
and $p/T_c<<1$ (in a continuum language) with $m_{glue}$
denoting the mass gap in the pure gauge sector. The small 
$\widehat{p}$-expansion
of the gauge field correlators induces an expansion of the action for 
the scalar fields with leading ultralocal and local
terms. Non-local terms are suppressed with higher momenta $\widehat{p}$.
The $\widehat{p}=0$ term of $W(\rho)$, $W^{(\circ)}$, 
is calculated in
a Monte Carlo simulation for a pure gauge theory in a constant
background field $r\equiv \rho(x)$ \cite{pinn}. 
The Monte Carlo results have been
compared with results from a small field and large field expansion of
$W$ in terms of $\rho$.
It turns out that the final value for the critical
coupling $\kappa_c$ sensitively depends on the shape of 
$W^{(\circ)}(\kappa r^2)$ for small values of $\kappa r^2$,
but is rather insensitive to the precise form for large arguments. 
Thus we improve on the small field range
by including contributions of the order $\widehat{p}^2$ to the 
first term in a
small field expansion, but neglect $\widehat{p}^2$ and higher order terms
of a large $\rho$-expansion, the leading conribution of which is absorbed
in $W^{\circ}(\kappa r^2)$.
The final form of the effective scalar model $S_{eff}(z)$ in terms of
scalar fields $z\equiv z(x)=\rho^2(x)$ is then given by
\be\label{seff}
S_{eff}(z) =\sum_{x \in \Lambda}
    \biggl( z + \lambda (z-1)^2 
                                      + W^{(\circ)}(\kappa z;\beta) 
                + \frac{D\kappa^2}{2} z^2  - 
               \frac{\kappa^2}{2} 
               \sum_{\mu} z(x) z(x+\widehat{\mu})
    \biggr) .
\ee
Although this model looks at most quadratic in $z$, it is non-Gaussian
because of the term $W^{(\circ})$ resulting from the Monte Carlo 
determination of the $\widehat{p}=0$-part of the effective action $W$
and $z\geq 0$. 
For given
$\beta$, $W^{(\circ)}=W^{(\circ)}(\kappa z)$ is known in the form of a
table for $\simeq 500$ values of $\kappa z\in [0,25]$. Note 
that the model is
no longer $Z(2)$ symmetric as the original SU(2) Higgs model was in 
terms of $\Phi$. The $Z(2)$ symmetry will only effectively be restored
at the critical endpoint. The missing Z(2) symmetry is also responsible
for the generalization of LCEs to LCEs in an "external field"
\cite{reisz2}. In terms of $z$ the hopping term has its usual form, but
it should be further noticed that now also two parts of the ultralocal part
of the action do depend on $\kappa$. It is possible to account for this
additional $\kappa$-dependence by a renormalization of the 
ultralocal vertices in the LCE graphs: the former constant coefficients in
the series for the susceptibilities now are expanded themselves
as power series in $\kappa$.
The phase structure of this model is next studied with
an effective potential with coefficients calculated to a high order in
the expansion parameter of the series.

\subsection{$\lambda_c$ from an Effective Potential with LCEs}

We approximate the partition function $Z(J)$ of the original SU(2) Higgs
model in 4D by $Z(J)=\int {\cal D} z\;\exp{-S_{eff}(z)+J\cdot z}$ . 
The effective
action $\Gamma(z)$ is then derived in the standard way by a 
Legendre transformation of $\ln{Z(J)}$ with $z\equiv d\ln{Z(J)/dJ}$.
Evaluating $\Gamma(z)$ for constant field configurations $z$ 
(corresponding
to constant currents $J$), we obtain the effective potential $V_{eff}$.

Next we express the coefficients of $V_{eff}$
in terms of quantities that are directly available in a high order linked 
cluster expansion. These are the susceptibilities that are 1PI in the
LCE-
graphical sense and are denoted by $\chi^{1PI}_n$.
The $\chi_n^{1PI}$, $n=2,\dots ,6$, are calculated up to and
including O($\kappa^{16}$).
If we expand the effective potential in fluctuations about the vacuum 
expectation value $\widetilde{z}:=d\ln{Z(J)}/dJ \vert_{J=0}$, 
the qualitative form looks like
\be\label{veff}
V_{eff}(x) = a_2 x^2 + a_3 x^3 +a_4 x^4 + O(x^5)
\ee
with $x = z-\widetilde{z}$ and coefficients $a_i$, $i=2,3,4$ that
depend on $\beta$, $\lambda$, $\kappa$ in a very implicit way. 
The phase
structure of the effective scalar model with action $S_{eff}$ is derived
by scanning the sign of $D:= (3a_3/8a_4)^2-(a_2/2a_4)$ as a function 
of $\kappa$. A first order transition is indicated by $D>0$ for a certain
range of $\kappa$s between the occurrence and disappearance of the
metastable second minimum in the symmetric and broken phase, 
respectively. In particular the critical endpoint $(\lambda_c,\kappa_c)$
of the first order transition line for fixed $\beta$ shows up as $D=0$
and $a_2=0=a_3$, or, in terms of the $\chi^{1PI}$s,
$4\kappa^2 \chi_2^{1PI}(\kappa,\lambda;\beta)=1$ and
$\chi_3^{1PI}(\kappa,\lambda;\beta)= 0$.
If we extrapolate the results for $\lambda_c(L)$, $\kappa_c(L)$, 
$L=7,.., 16$, to infinite order of truncation,
$\lambda_c(L) = \lambda_c(\infty)\;+\;const\; 1/L$, we obtain
$\lambda_c(\infty)\; =\; 0.0032(1)$
and, analogously, $\kappa_c(\infty)= 0.1447(1)$ .
An inclusion of $\chi^{1PI}_5$ and $\chi^{1PI}_6$-terms in the 
effective potential
shows that the neglected coefficients in (\ref{veff}) are suppressed
by more than an order of magnitude.



\section{Outlook}

Feasible extensions in
our framework include a calculation for the gauge coupling $\beta=10$,
larger extensions in time or larger values of 
$\lambda$.
So far the LCE-programs were run on a SUN-workstation.
Applied to the deconfinement transition of QCD with dynamical fermions, 
the first approach 
would lead to a localization of the line of critical hopping
parameters, whereas a derivation of an effective scalar model
(in terms of quark condensates) along the lines of the second approach
applies to the chiral transition of QCD.
Finally we remark that DLCEs 
have a much wider range of 
applications than it was indicated here.
They include spin glasses,
partially annealed neural networks, or diluted Ising models and are 
independent of any variational approach.

\end{document}